\documentclass[amssymb,amsmath,floatfix,prl,twocolumn,superscriptaddress,nolongbibliography]{revtex4-2}
\usepackage[latin1]{inputenc}

\usepackage{mathptmx}
\usepackage[T1]{fontenc}
\usepackage{graphicx}
\usepackage{amsmath,amssymb,amsfonts, MnSymbol}
\usepackage{mathrsfs}
\usepackage{bm}
\usepackage{hyperref}
\usepackage{bbold}
\usepackage{xcolor}

\newcommand{\algn}[1]{\begin{align} #1 \end{align}}

\newcommand{\ee}{\ensuremath{\text{e}}}
\newcommand{\ed}{\ensuremath{\text{d}}}

\newcommand{\ms}[1]{\ensuremath{\mathscr{#1}}}

\newcommand{\kb}{\ensuremath{k_\text{B}}}
\newcommand{\eqnlab}[1]{\label{eq:#1}}

\newcommand{\figlab}[1]{\label{fig:#1}}
\newcommand{\eqnref}[1]{\eqref{eq:#1}}
\newcommand{\Eqnref}[1]{Eq.~\eqref{eq:#1}}
\newcommand{\Eqsref}[1]{Eqs.~\eqref{eq:#1}}

\newcommand{\figref}[1]{\ref{fig:#1}}
\newcommand{\Figref}[1]{Fig.~\ref{fig:#1}}
\newcommand{\Figsref}[1]{Figs.~\ref{fig:#1}}

\begin{document}
\title{Finite-Time Dynamical Phase Transition in Non-Equilibrium Relaxation}

\author{Jan Meibohm}
\affiliation{Complex Systems and Statistical Mechanics, Department of Physics and Materials Science, University of Luxembourg, L-1511 Luxembourg, Luxembourg}
\author{Massimiliano Esposito}
\affiliation{Complex Systems and Statistical Mechanics, Department of Physics and Materials Science, University of Luxembourg, L-1511 Luxembourg, Luxembourg}

\begin{abstract}
We uncover a finite-time dynamical phase transition in the thermal relaxation of a mean-field magnetic model. The phase transition manifests itself as a cusp singularity in the probability distribution of the magnetisation that forms at a critical time. The transition is due to a sudden switch in the dynamics, characterised by a dynamical order parameter. We derive a dynamical Landau theory for the transition that applies to a range of systems with scalar, parity-invariant order parameters. Close to criticalilty, our theory reveals an exact mapping between the dynamical and equilibrium phase transitions of the magnetic model, and implies critical exponents of mean-field type. We argue that interactions between nearby saddle points, neglected at the mean-field level, may lead to critical, spatiotemporal fluctuations of the order parameter, and thus give rise to novel, dynamical critical phenomena.
\end{abstract}
\maketitle
The dynamic response of many-body systems to changes of the external parameters, be it the temperature, pressure or an external field, is of fundamental interest in statistical mechanics and has a wide range of applications. When the changes are small, the response of the system is linear~\cite{Ons31a,Ons31b,Kub57a,Kub57b}, and rather well understood~\cite{Che09,Fal16,Bai13}. In many applications, however, the external parameters switch suddenly and violently, thus driving the system far away from equilibrium. Non-equilibrium relaxation phenomena are theoretically~\cite{Lif62,Pal84,Cug03} and experimentally~\cite{Wee02,Wan06,Far13} challenging, in particular when they exhibit long transients~\cite{Bra02,Ber11}, which is often the case in the presence of phase transitions.

Equilibrium phase transitions are qualitative changes of a system's equilibrium state under adiabatic variation of the external parameters~\cite{Cal85}. They are accompanied by characteristic changes of order parameters~\cite{Cha95}, such as the density or the magnetisation. At continuous phase transitions, thermodynamic quantities and order parameters exhibit power-law behaviour~\cite{Gol92}, and the values of their critical exponents divide systems into universality classes.  Recent developments~\cite{Fre84,Gra84,Ell07,Sei12,Ber15,Pel21} have given rise to conceptual generalisations of phase transitions to non-equilibrium systems~\cite{Der87,Ge11,Tom12,Her18,Vro20,Mar20} and dynamic observables~\cite{Meh08,Lac08,Jac10,Nya16,Nem19,Laz19,Sun19,Her20}. These ``dynamical phase transitions'' are related to qualitative changes in the \textit{dynamics}~\cite{Gar07,Nya18,Jac20,Pro20,Ket21}, observed in the long-time limit, under varying external conditions.

In this Letter, we uncover a \textit{finite-time} dynamical phase transition in the non-equilibrium relaxation of a classical, mean-field spin model. In distinction to other classical transitions, the present one occurs in the transient response to an instantaneous change (a ``quench'') of the temperature that induces an order-to-disorder phase transition. Interestingly, dynamical phase transitions with similar properties have recently been found in closed quantum systems~\cite{Hey13,Hey18}. The transition manifests itself as a transient cusp singularity in the probability distribution of the magnetisation and is the result of competing dynamic behaviours within the system. The interpretation of this cusp as a phase transition sheds new light on previous works~\cite{Ent02,Kul07,Ent10,Erm10,Red12,Fer13} that discuss mathematical details of the singularity, and the existence and absence of a Gibbs measure for the transient. 
We derive a dynamical Landau theory for the phase transition that is robust against symmetry preserving transformations, and which applies to a range of systems with scalar, parity-invariant order parameters. An exact mapping between the dynamical and equilibrium phase transitions of the magnetic model classifies the transition as continuous, with mean-field-type critical exponents.

The Curie-Weiss model consists of $N\to\infty$ coupled Ising spins $\sigma_i=\pm1$, $i=1,\ldots,N$, with infinite-range, ferromagnetic interaction of strength $J/(2N)$. The system is immersed in a heat bath at inverse temperature $\beta=1/(\kb T)$, and subject to an external field $H$. Because of the mean-field nature of the interaction, all states with equal numbers $N_{\pm}$ of spins in the states $\pm1$, respectively, are equivalent. Therefore, any microstate can be written in terms of the total magnetisation $M = N_+-N_-$. The free energy $F$ reads~\cite{Cha95}
\algn{\eqnlab{epot}
 	F(M) = -\frac{J}{2N}\left(M^2 -N\right)-M H - \beta^{-1}S(M)\,,
}
where the dimensionless internal entropy $S(M) = \ln\Omega(M)$ originates from the microscopic degeneracy $\Omega(M)$ of $M$.

We endow the system with a stochastic dynamics mediated by thermal fluctuations of the heat bath. A transition $\mp1\to \pm1$ of an arbitrary spin leads to $M\to M_\pm\equiv M\pm 2$. The evolution of the probability $P(M,t)$ for finding the system in state $M$ at time $t$ is described by the master equation
\algn{\eqnlab{mastereqn}
	\dot P(M,t)=\sum_{\eta=\pm}\left[W_\eta(M_{-\eta})P(M_{-\eta},t)-W_\eta(M)P(M,t)\right]\,,
}
where $W_\pm(M)$ are the rates for $M\to M_\pm$, given by
\algn{\eqnlab{rates}
	W_\pm(M) =\frac{N\mp M		}{2\tau}\exp\left\{ \pm\beta\left[\frac{J}{N}\left(M \pm 1	\right)	+ H\right]\right\}\,,
}
with microscopic relaxation time $\tau$. The algebraic prefactor $\propto (N\mp M)/2= N_{\mp}$ is due to equivalence of microscopic transitions, and spin-inversion invariance implies parity symmetry in $M$, $W_\pm(M)|_{H=0}=W_\mp(-M)|_{H=0}$. Forward and backward rates are related by detailed balance~\cite{Kam07}, $W_\pm(M)P^\text{eq}(M)=W_\mp(M_\pm)P^\text{eq}(M_\pm)$, with respect to the equilibrium distribution $P^\text{eq}(M) =  Z^{-1}\exp\left[-\beta F(M)\right]$; $Z$ denotes the partition function.
\begin{figure}
	\centering
	\includegraphics[width=\linewidth]{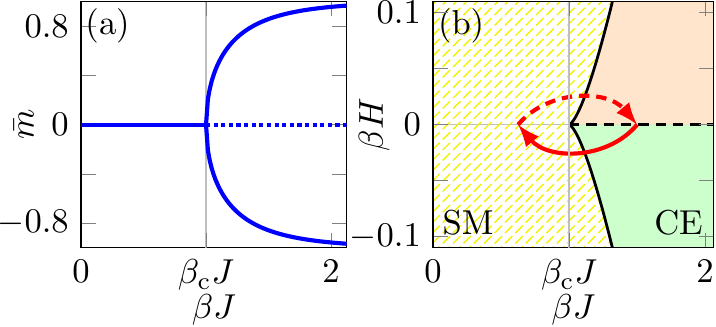}
	\caption{(a) Magnetisation $\bar m$ at $H=0$. Solid lines show minima of $\ms{V}^\text{eq}$, the dotted line a local maximum. (b) Equilibrium phase diagram, featuring the SM phase (lined) and CE phase (filled), separated by the phase boundary (solid black line). At the dashed line, $\bar m$ jumps discontinuously. Red arrows indicate disordering (solid line) and ordering (dashed line) quenches, respectively. }\figlab{phase_diagram}
\end{figure}

To take the thermodynamic limit, $N\to\infty$, we define the intensive magnetisation $m\equiv M/N$ per spin and the free-energy density $\ms{F}(m) \equiv F(Nm)/N$. The equilibrium distribution for $m$ then takes the large-deviation form $P^\text{eq}(m) \smilefrown \exp\left[-N \ms{V}^\text{eq}(m)\right]$~\cite{Ell07,Tou09} with equilibrium rate function $\ms{V}^\text{eq}(m)=\beta\left[\ms{F}(m)-\ms{F}(\bar m)\right]$, where $\ms{F}(m)= -Jm^2/2-m H - \beta^{-1} \ms{S}(m)$ and $\ms{S}(m)=-(1/2)\sum_{\eta=\pm} \left(1+\eta m\right)\ln\left(1+\eta m\right)+\ln2$. The term $\ms{F}(\bar m)= \min_m \ms{F}(m)$ originates from the normalisation of $P^\text{eq}(M)$ and $\bar m$ denotes the order parameter, the mean magnetisation $\bar m \approx \langle m \rangle$ in the thermodynamic limit. Expanding $\ms{V}^\text{eq}(m)$ to quartic order in $m$, one finds
\algn{\eqnlab{eqrf}
	\ms{V}^\text{eq}(m) \sim -\beta  H m - J\left(\beta-\beta_\text{c}\right)\frac{m^2}2+\frac{m^4}{12}-\beta\ms{F}(\bar m)\,,
}
where the quadratic term changes sign at $\beta_\text{c} = 1/J$ while the quartic term remains positive. Hence, for $H=0$, $\ms{V}^\text{eq}(m)$ passes from a single well to a symmetric double well at the critical inverse temperature $\beta_\text{c}$. This corresponds to a continuous phase transition~\cite{Lan37} from a disordered into an ordered state, that spontaneously breaks the parity symmetry in $m$.

Close to $\beta_\text{c}$, $\bar m$ changes continuously from $\bar m =0$ to finite $\bar m$, as shown in \Figref{phase_diagram}(a). Figure~\figref{phase_diagram}(b) shows the phase diagram of the model, exhibiting two distinct phases, separated by a phase boundary (solid black line): a single-mode (SM) phase, where $\ms{V}^\text{eq}$ has a unique minimum, and a coexistence (CE) phase, where local and global minima coexist. Within the CE phase, $H\neq0$ lifts the degeneracy between the two minima in $\ms{V}^\text{eq}(m)$, and thus splits the CE phase into regions with $\bar m>0$ (orange) and $\bar m<0$ (green). Moving across the dashed line that separates the two, $\bar m$ jumps discontinuously to $-\bar m$.

\begin{figure*}
	\includegraphics[width=\linewidth]{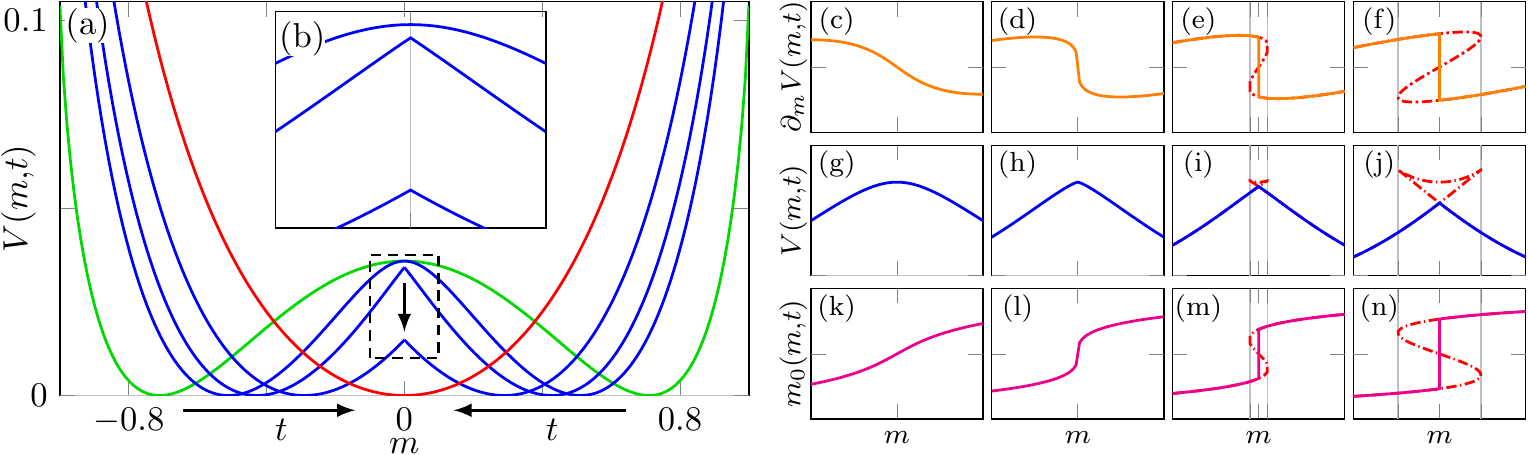}
	\caption{Post-quench dynamics for $\beta=5/(4J)$ and $\beta_\text{q}=3/(4J)$. (a) Rate functions $\ms{V}^\text{eq}$ (green), $\ms{V}^\text{eq}_q$ (red), and $V(m,t)$ for $t/\tau=0.5$, $0.8$ and $1.5$ (blue). Black arrows indicate the time evolution of $V(m,t)$. (b) Magnification of the dashed rectangle in \Figref{quench}(a). (c)--(n) Time evolution of $\partial_mV(m,t)$ [(c)--(f)], $V(m,t)$ [(g)--(j)] and $m_0(m,t)$ [(k)--(n)] for $t/\tau=0.5$, $0.7$, $0.8$, and $1$ (from left to right). The solid and dash-dotted lines show dominant and sub-dominant solutions, respectively (see main text). Gray lines delimit the dynamical coexistence region.}\figlab{quench}
\end{figure*}

In the vicinity of the critical point, $(\beta,H)=(\beta_\text{c},0)$, we obtain from the minimisation condition $\ed \ms{V}^\text{eq}/\ed m|_{m=\bar m}=0$, an equation of state that leads to mean-field critical exponents~\cite{Gol92,Cha95}, universal among mean-field models. In particular, for $H=0$ one finds that $\bar m(\beta)$ is continuous, with $\bar m=0$ and $\bar m \propto |\beta-\beta_\text{c}|^{1/2}$, below and slightly above $\beta_\text{c}$, respectively.

Starting from an ordered equilibrium state in the CE phase with $\beta>\beta_\text{c}$ at time $t<0$, we induce an instantaneous disordering quench $\beta\to\beta_\text{q}$ into the SM phase, $\beta_\text{q}<\beta_\text{c}$, at $t=0$. For simplicity, we set $H=0$. Because the quench crosses the phase boundary [solid arrow in \Figref{phase_diagram}(b)], it induces an order-to-disorder phase transition. In contrast to ordering quenches [dashed arrow in \Figref{phase_diagram}(b)]~\cite{Bra02}, disordering quenches are ergodic, so that $P(m,t)\to P^\text{eq}_q(m)\smilefrown\exp[-N\ms{V}^\text{eq}_q(m)]$ as $t\to\infty$, where $\ms{V}^\text{eq}_q(m)$ is the equilibrium rate function at the final inverse temperature $\beta_\text{q}$.

For $t>0$, the post-quench dynamics of the probability distribution $P(m,t)\smilefrown\exp[-NV(m,t)]$, with time-dependent rate function $V(m,t)$, is determined by \Eqnref{mastereqn} which turns into the Hamilton-Jacobi equation
\algn{\eqnlab{hj}
	0=\partial_t V(m,t) + \ms{H}[m,\partial_m V(m,t)]\,,
}
to leading order in $N\gg1$, with initial condition $V(m,0)=\ms{V}^\text{eq}(m)$ and Hamiltonian~\cite{Dyk94,Imp05,Fen06}
\algn{
	\ms{H}(q,p) &= w_+(q)\left(\ee^{2p}-1\right) + w_-(q)\left(\ee^{-2p}-1\right)\,;
}
see Sec.~I in the Supplemental Material~\cite{supp}. Here, $w_\pm(q) = (2\tau)^{-1}(1\mp q)\exp\left(\pm \beta_\text{q} J q \right)$ denote the $N$-scaled transition rates. Solutions to \Eqnref{hj} are given in terms of characteristics $q(s)$ and $p(s)$, $0\leq s\leq t$, that solve the Hamilton equations~\cite{Cou62}
\algn{\eqnlab{heom}
	\dot q(s) = \partial_p \ms{H}(q,p)\,,\quad \dot p(s) = -\partial_q \ms{H}(q,p)\,,
}
with boundary conditions
\algn{\eqnlab{bound}
	p(0) = \tfrac{\ed}{\ed m} \ms{V}^\text{eq}[q(0)]\,, \qquad q(t) = m\,.
}
From the solutions of \Eqsref{heom}--\eqnref{bound}, $V(m,t)$ is obtained as
\algn{\eqnlab{vft}
	V(m,t)  =  \int_0^t \ed s \left[ p \dot q  - \ms{H}(q,p) \right] + \ms{V}^\text{eq}[q(0)]\,.
}
Any solution of \Eqsref{heom} and \eqnref{bound} solves the variational problem $\delta V(m,t)=0$, where $\delta$ denotes a variation over all trajectories with final point $q(t)=m$~\cite{Gol80,Cou62}, i.e., all characteristics $[q(s),p(s)]_{0\leq s\leq t}$ are saddle points of \Eqnref{vft}. From the large-deviation form $P(m,t)\smilefrown \exp[-NV(m,t)]$ we conclude that only those characteristics that minimise $V(m,t)$ contribute for $N\to\infty$, and the others are exponentially suppressed. The minimising characteristic $q(s)$, $0\leq s\leq t$, constitutes the system's optimal fluctuation, the most likely way to realise the magnetisation $q(t)=m=\lim_{N\to\infty}M(t)/N$ at time $t$, in response to the quench at $t=0$. In particular, $m_0(m,t)\equiv q(0)$ corresponds to the most likely \textit{initial} magnetisation, which plays the role of an order parameter, as we explain below.

We compute $V(m,t)$ by solving \Eqsref{heom}--\eqnref{bound} with a shooting method~\cite{Pre86} on a fine $m$ grid, see Sec.~II in the Supplemental Material. We extract three fields: $V(m,t)$ [by evaluating \Eqnref{vft}], the derivative field $\partial_mV(m,t) = p(t)$ [the end point of $p(s)$], and the initial magnetisation $m_0(m,t)=q(0)$.

The blue curves in \Figref{quench}(a) show $V(m,t)$ for different $t$; the green and red curves show $\ms{V}^\text{eq}$ and $\ms{V}^\text{eq}_q$, respectively. For small times, $V(m,t)$ is a symmetric double well, similar to the initial $\ms{V}^\text{eq}$. As $t$ increases, the minima of $V(m,t)$ move towards the origin, and the local maximum at $m=0$ decreases, as indicated by the black arrows in \Figref{quench}(a). In the long-time limit, $V(m,t)$ approaches the single-mode shape of $\ms{V}^\text{eq}_q$. 

Notably, however, $V(m,t)$ does not evolve smoothly: At a finite time $t$, $V(m,t)$ forms a cusp at $m=0$ [see \Figref{quench}(b)], and the derivative field $\partial_mV(m,t)$ develops a discontinuous jump. The origin of this jump is shown in \Figsref{quench}(c)--\figref{quench}(f): As time evolves, $\partial_mV(m,t)$ folds over and becomes multi-valued, and up to three solutions of \Eqsref{heom} and \eqnref{bound} coexist within a finite interval, delimited by the gray lines in \Figsref{quench}(e) and \figref{quench}(f). Selecting the one with smallest $V(m,t)$ naturally leads to a Maxwell construction for the dominant solution, shown in orange. The sub-dominant solutions [red, dash-dotted lines in \Figref{quench}(c)--\figref{quench}(f)] have a larger $V(m,t)$ as shown in \Figsref{quench}(g)--(j). Figures~\figref{quench}(k)--\figref{quench}(n) indicate the same multivaluedness, and an analogous Maxwell construction for $m_0(m,t)$, culminating in a discontinuous jump at $m=0$.

Interpreting the formation of the cusp as a finite-time dynamical phase transition, we exploit the similarities with the equilibrium transition of the model. We first note that the sudden change of $m_0(m,t)$ at $m=0$ is strikingly similar to the discontinuous jump of $\bar m$ at equilibrium, when the external field $H$ crosses zero in the CE phase [dashed line in \Figref{phase_diagram}(b)]. To be specific, we identify $t$, $m_0$, and $m$ in the dynamical case with $\beta$, $\bar m$, and $H$, respectively, at equilibrium, and draw a ``dynamical phase diagram'', shown in \Figref{phase_diagram_tx}(a): Small times $t<t_\text{c}$ correspond to the dynamical single-mode (DSM) phase (yellow, lined region) where the dynamical order parameter $m_0(m,t)$ is unique (just like $\bar m$ for $\beta<\beta_\text{c}$) and $V(m,t)$ is a smooth function of $m$. For $t>t_\text{c}$ the system transitions into a dynamical coexistence (DCE) phase (filled region) where multiple $m_0$ values coexist. The DCE phase corresponds to the $m$ interval delimited by the gray lines in \Figsref{quench}(m) and \figref{quench}(n). For $m=0$, the two values, $m_0$ and $-m_0$ are degenerate, and parity symmetry is spontaneously broken by the dynamics, in analogy with $\bar m$ and $-\bar m$ for $\beta>\beta_\text{c}$ and $H=0$. Conditioning on $m\neq0$ lifts this degeneracy, so that one of the $m_0$ values becomes exponentially suppressed. Consequently, when crossing the dashed line in the DCE phase in \Figref{phase_diagram_tx}(a), $m_0$ jumps discontinuously, leading to the kink in $V(m,t)$ at $m=0$.
\begin{figure*}
	\includegraphics[width=\linewidth]{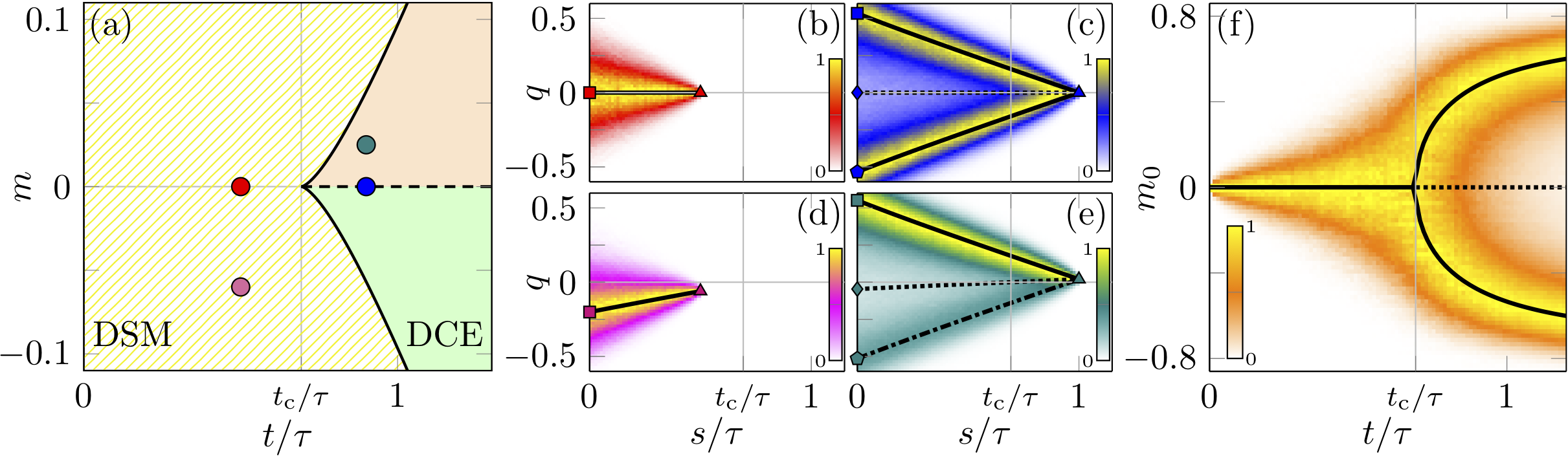}
	\caption{(a) Dynamical phase diagram for $(\beta,\beta_\text{q})=[5/(4J),3/(4J)]$, featuring the DSM (yellow, lined) and DCE (filled) phases (see main text), separated by a phase boundary (solid black line). The DCE phase splits into regions with $m_0(m,t)>0$ (orange) and $m_0(m,t)<0$ (green). At the dashed line, $m_0(m,t)$ jumps discontinuously. The bullets show $(t,m)$ values of the equally coloured trajectory densities in \Figref{phase_diagram_tx}(b)--\figref{phase_diagram_tx}(e). (b)--(e) Optimal fluctuations from theory (solid black line) and from numerical simulations (coloured regions, $N=200$, $10^8$ trajectories). Sub-dominant trajectories are shown as broken lines, symbols mark $m_0(m,t)=q(0)$ and $m=q(t)$. (f) Density of $m_0$ at $m=0$ as function of $t$ from numerical simulations (coloured region, $N=200$, $10^8$ trajectories), and optimal (solid black line) and sub-dominant (dotted line) $m_0(0,t)$ from theory.}
	\figlab{phase_diagram_tx}
\end{figure*}

On the trajectory level, the transition from the DSM into the DCE phase corresponds to a sudden change of the optimal fluctuation that minimises $V(m,t)$ in \Eqnref{vft}~\cite{Ent10,Erm10}. This follows from the dynamical analogue of an energy-entropy argument~\cite{Gol92}: For small times, the right-hand side of \Eqnref{vft} is dominated by the first term, interpreted as an activity. For $m=0$, this term is minimised by the inactive solution $p(s)=q(s)=0$, but at the cost of an unfavourable initial $q(0)=0$, the local maximum of the static, second term $\ms{V}^\text{eq}$. For long times, the activity term loses its dominance and $\ms{V}^\text{eq}$ becomes important so that the optimal fluctuation optimises $q(0)$ by starting at the minima of $\ms{V}^\text{eq}$.

To visualise the optimal fluctuations, we perform numerical simulations at finite $N$. From \Eqnref{mastereqn}, we generate a large number of trajectories and record $q(s)\approx M(s)/N$, for different $0\leq s \leq t$, conditioning them to end up at $q(t) \in [m-\ed m,m+\ed m]$ for given $t$ and $m$, and a small $\ed m$.  We then collect histograms of $q(s)$, normalised to unity for each $s$, and merge them, to obtain a numerical approximation of the trajectory density, shown in \Figsref{phase_diagram_tx}(b)--\figref{phase_diagram_tx}(e). The colour coding matches their $(t,m)$ values, shown as the equally-coloured bullets in \Figref{phase_diagram_tx}(a).

Figures~\figref{phase_diagram_tx}(b) and \figref{phase_diagram_tx}(c) show good agreement between the optimal fluctuations (solid black line) and the yellow regions of high trajectory density in the DSM and DCE phase, respectively, at $m=0$. In the DSM phase [\Figref{phase_diagram_tx}(b)], we observe a unique optimal fluctuation that remains at zero, the inactive solution. Beyond the dynamical critical point [\Figref{phase_diagram_tx}(c)], two degenerate optimal fluctuations coexist, with initial magnetisations $m_0$ and $-m_0$ close to the minima of $\ms{V}^\text{eq}$. The third trajectory (dotted line) has a larger $V(m,t)$, and is not observed in the numerics. 

Figures~\figref{phase_diagram_tx}(d) and \figref{phase_diagram_tx}(e) show the optimal fluctuations and trajectory densities for finite $m$. In the DSM phase [\Figref{phase_diagram_tx}(d)], the conditioning shifts the optimal fluctuation away from zero. Finite $m$ in the DCE phase [\Figref{phase_diagram_tx}(e)] lifts the degeneracy between the optimal fluctuations, so that one of them is degraded to a local minimum of $V(m,t)$ (dash-dotted line), whose remnants persist at finite $N$.

In \Figref{phase_diagram_tx}(f) we visualise the dependence of the order parameter $m_0(m,t)$ on $t$ by joining the histograms of $m_0=q(0)\approx M(0)/N$ for $m=0$ and different $t$. The solid black line shows the theoretical prediction for $m_0(0,t)$, obtained from \Eqsref{heom}--\eqnref{bound}. Apart from the excellent agreement between the yellow regions and the theoretical curves, we observe a close similarity with the dependence of $\bar m$ on $\beta$, see \Figref{phase_diagram}(a).

To classify the dynamical phase transition in terms of equilibrium categories, we express $V(m,t)$ as the minimum of a dynamical Landau potential $\ms{L}^{m,t}(m_0)$, $V(m,t)=\min_{m_0}\ms{L}^{m,t}(m_0)-\beta\ms{F}(\bar m)$, with the minimiser given by the dynamical order parameter $m_0(m,t)$. We then show that this provides an exact mapping between the dynamical phase diagram and the equilibrium one, close to the critical point.

First, we calculate the critical time $t_\text{c}$ for the transition. Since $V(0,t)$ develops a kink at time $t=t_\text{c}$, the curvature $z(t)\equiv\partial_m^2V(0,t)$ must tend to negative infinity, $z(t)\to-\infty$ as $t\to t_\text{c}$. Taking a partial derivative of \Eqnref{heom}, we find that $z(t)$ obeys a Riccati equation whose solution tends to $-\infty$ at $t_\text{c}/\tau = \ln \left[(\beta-\beta_\text{q})/(\beta-\beta_\text{c})\right]/[4(\beta_\text{c}-\beta_\text{q})J]$, see Sec.~III in the Supplemental Material for details, and Ref.~\cite{Erm10} for a different method. With the parameters of \Figsref{quench} and \figref{phase_diagram_tx}, we obtain $t_\text{c}/\tau=\ln(2)\approx 0.6931$, in agreement with the numerical result.

We now derive the Landau potential $\ms{L}^{m,t}(m_0)$, whose minimum is attained by $m_0(m,t)$, analogous to the minimum $\bar m$ of $\ms{V}^\text{eq}$ at equilibrium. Close to the critical point, we expand $\ms{H}$ perturbatively to fourth order in $q$ and $p$, $\ms{H}\sim \ms{H}_0 + \ms{H}_1$. A sequence of canonical transformations brings the quadratic and quartic Hamiltonians $\ms{H}_0$ and $\ms{H}_1$ into the forms $\ms{H}_0=(p^2-q^2)/2$ and $\ms{H}_1 = \alpha(p^4+q^4)/4$, with real parameter $\alpha(\beta,\beta_\text{q},J)$. Using \Eqnref{vft}, we then compute $\ms{L}^{m,t}(m_0)$ perturbatively to fourth order in $m_0$ and lowest order in $m$, see Sec.~IV in the Supplemental Material. We find
\algn{\eqnlab{dlp}
	\ms{L}^{m,t}(m_0) \sim -m\lambda_0m_0 - \lambda_1\left(\frac{t-t_\text{c}}\tau\right) \frac{m_0^2}4 + \left[\lambda_2 + \ldots	\right]\frac{m_0^4}4\,,
}
where $\lambda_{0,1}>0$ and $\lambda_2>0$ for $\beta<3/(2J)$, and arbitrary $\beta_\text{q}$ and $J$; the dots denote higher-order terms in $t-t_\text{c}$. For $m=0$, $\ms{L}^{m,t}(m_0)$ transitions from a single into a double well at the critical time $t_\text{c}$, just like $\ms{V}^\text{eq}$ at $\beta_\text{c}$. In other words, \Eqnref{dlp} is the dynamical analogue of \Eqnref{eqrf} and both expressions can be mapped onto each other by proper rescaling of, e.g., $t-t_\text{c}$, $m_0$, and $m$. Hence, the dynamical phase transition has mean-field critical exponents and $m_0$ changes continuously from $m_0=0$ to $m_0\propto|t-t_\text{c}|^{1/2}$, below and above $t_\text{c}$, respectively. By contrast, for $\beta>3/(2J)$ we can have $\lambda_2<0$, and the system may undergo a discontinuous, first-order dynamical phase transition, where $m_0(m,t)$ jumps discontinuously. Symmetry-preserving transformations of the rates $W_\pm(M)\to\tilde W_\pm(M)$ leave $\lambda_{0,1}$ invariant, suggesting that the occurrence of the transition is model independent (Sec.~V in the Supplemental Material).

Our dynamical Landau theory applies to an entire class of systems with scalar, parity-symmetric order parameters. A particularly simple one, that could be realised in a state-of-the-art experiment~\cite{Cil17,Kum20}, is a Brownian particle at weak noise, subject to a potential quench from a double into a single well  (Sec.~IV.B in the Supplemental Material). 

Close to the critical point, interactions between nearby saddles $\delta V(m,t)=0$ of \Eqnref{vft}, neglected here, may lead to critical, spatiotemporal fluctuations of the order parameter, analogous to equilibrium~\cite{Gol92,Cha95}, and give rise to corrections to the mean-field exponents. The presence of strong fluctuations is indicated by the divergence $\propto |t-t_\text{c}|^{-1/2}$ of the pre-exponential factor of $P(m,t)$ at the critical point, see Sec.~VI in the Supplemental Material. Since these fluctuations are of dynamical origin, we hypothesise that their effects are different from the equilibrium ones, and thus reflect a novel, dynamical critical phenomenon. This can be tested by investigating the post-quench dynamics of systems with short-range interactions in two and three dimensions using Monte-Carlo simulations~\cite{Lan14}, or perhaps dynamic renormalisation group methods~\cite{Tau14}.

\begin{acknowledgments}
	We thank Gianmaria Falasco and Nahuel Freitas for discussions, and Karel Proesmans for pointing out the connection with the mathematics literature. This work was supported by the European Research Council, project NanoThermo (ERC-2015-CoG Agreement No. 681456).
\end{acknowledgments}
\end{document}


\title{Supplemental Material for: Finite-Time Dynamical Phase Transition in Non-Equilibrium Relaxation}

\author{Jan Meibohm}
\affiliation{Complex Systems and Statistical Mechanics, Department of Physics and Materials Science, University of Luxembourg, L-1511 Luxembourg, Luxembourg}
\author{Massimiliano Esposito}
\affiliation{Complex Systems and Statistical Mechanics, Department of Physics and Materials Science, University of Luxembourg, L-1511 Luxembourg, Luxembourg}
\begin{abstract}
In this Supplemental Material, we give additional details on the derivations and the numerical method discussed in the main text. We first derive the Hamilton-Jacobi equation [Eq.~(5) in the main text] and explain the shooting method we use to solve it. From the Hamilton equations we then derive the Riccati equation for the curvature $z(t)$. We then show how to obtain the dynamical Landau potential [Eq.~(10) in the main text] for a general, fourth-order dynamical Hamiltonian, by means of canonical transformations. The final expression for the dynamical Landau potential depends only on five parameters and one time scale. We specify these parameters for the Curie-Weiss model and for a Brownian particle subject to a potential quench, to prove the occurrence of the dynamical phase transition for these systems. We use the general form of the dynamical Landau potential to show that the phase transition is robust against symmetry-preserving transformations of the transition rates in the Curie-Weiss model. Finally, we show that the pre-exponential factor associated with the large-deviation form of $P(m,t)$ diverges at the critical point, indicating the occurrence of strong, and perhaps critical, fluctuations.
\end{abstract}
\maketitle
\section{Derivation of Hamilton-Jacobi equation}
In this section, we derive the Hamilton-Jacobi equation for $V(m,t)$, Eq.~(5) in the main text, from the Master equation for $P(M,t)$, Eq.~(2) in the main text:
\algn{\eqnlab{mastereqn}
	\dot P(M,t)=\sum_{\eta=\pm}\left[W_\eta(M_{-\eta})P(M_{-\eta},t)-W_\eta(M)P(M,t)\right],\quad\text{with rates}\quad W_\pm(M) =\frac{N\mp M		}{2\tau}\exp\left\{ \pm\beta_\text{q}\left[\frac{J}{N}\left(M \pm 1	\right)	+ H\right]\right\}\,,
}
where $M_\pm = M\pm2$. We first express $P(M,t)$ in terms of $V(m,t)$ as $P(M,t)=\exp[-NV(m,t)]$. Substituting this into the Master equation in \eqnref{mastereqn}, we obtain
\algn{\eqnlab{mastereqn2}
	-\partial_t V(m,t) = \ee^{\beta_\text{q}J/N}\sum_{\eta=\pm}\left\{w_{\eta}\left(m-\eta\frac2N\right)\exp\left\{N\left[V(m,t)-V\left(m-\eta\frac2N,t\right)\right]\right\}	- w_{\eta}(m)	\right\}\,,
}
with $w_{\pm}(m) = (1\mp m		)\exp\left[ \pm(\beta_\text{q}Jm	+ H)\right]/(2\tau)$.
For $N\gg1$, we find
\algn{\eqnlab{limit1}
	\exp\left\{N\left[V(m,t)-V\left(m\mp\frac2N,t\right)\right]\right\}\sim\exp\left[ \pm2\partial_m V(m,t)	\right]\left[1-\frac2N\partial^2_m V(m,t)\right]\,,
}
and
\algn{\eqnlab{limit2}
	 w_\pm\left(m\mp \frac2N\right) \sim w_{\pm}(m)\mp \frac2N\partial_m w_\pm(m)\,.
}
Using these asymptotic expressions, we obtain from \Eqnref{mastereqn2} to order $\mathscr{O}(1)$ in $N$,
\algn{\eqnlab{hjeqn}
	-\partial_t V(m,t) = \sum_{\eta=\pm}w_{\eta}(m)\left\{\exp\left[\eta 2\partial_mV\left(m,t\right)\right]	- 1\right\}\,,
}
i.e., Eq.~(5) in the main text. Equation~\eqnref{hjeqn} is a first-order partial differential equation for $V(m,t)$ which does not depend explicitly on $V(m,t)$ itself, but only on its partial derivatives. This characterises \Eqnref{hjeqn} as a Hamilton-Jacobi equation~\cite{Cou62} with Hamiltonian $\ms{H}(q,p)$ given in Eq.~(6) in the main text. The characteristics of \Eqnref{hjeqn} solve the Hamilton equations, Eq.~(7) in the main text. From the conjugate momentum $p(s) = \partial_mV[q(s),s]$ one obtains the boundary condition $p(0) = \partial_m V[q(0),0] = \ed\ms{V}^\text{eq}[q(0)]/\ed m$. Furthermore, we have
\algn{
	\tfrac{\ed}{\ed s} V[q(s),s] = p(s) \dot q(s) + \partial_s V[q(s),s] =  p(s) \dot q(s) - \ms{H}[q(s),p(s)]\,,
}
where we used the Hamilton-Jacobi equation in the second step. Integrating $s$ from $0$ to $t$ we readily obtain Eq.~(9) in the main text.
\section{Numerical solution of Hamilton-Jacobi equation}
In this section, we provide details on the numerical method used to calculate $V(m,t)$, $\partial_mV(m,t)$ and $m_0(m,t)$, shown in Fig.~2 of the main text. The Hamilton equations with boundary conditions, 
\algn{\eqnlab{bvp}
	\dot q(s) = \partial_p \ms{H}(q,p)\,,\quad \dot p(s) = -\partial_q \ms{H}(q,p)\,,\qquad p(0) = \tfrac{\ed}{\ed m}\ms{V}^\text{eq}[q(0)]\,,\qquad	q(t) = m\,,
}
Eqs.~(7) and (8) in the main text, consititute a two-point boundary value problem~\cite{Pre86} on the finite interval $q\in [-1,1]$. Problems of this kind can be solved to high accuracy by shooting~\cite{Pre86}. 

The basic idea is the following: We numerically integrate the Hamilton equations in \Eqsref{bvp} for a grid of initial values
\algn{
	q^{(i)}(0)=m_0^{(i)} \in [-1,1]\,,\quad \text{and}\quad p^{(i)}(0)=\tfrac{\ed}{\ed m}\ms{V}^\text{eq}\left[m_0^{(i)}\right]\,, \quad i = 1,\ldots M\,,
}
over given finite time $t$. For each grid index $i$, we obtain a trajectory $[q^{(i)}(s),p^{(i)}(s)]_{0\leq s\leq t}$ with end point $[q^{(i)}(t),p^{(i)}(t)]$. For a given $m$, the value $\eps^{(i)}=|q^{(i)}(t)-m|$ defines the error of the initial condition $m^{(i)}_0$. We use an interpolation algorithm to compute the function $\eps(m_0)$, of which we numerically determine the roots $m^*_0$, $\eps(m^*_0)=0$. Now, we use $q(0)=m_0^*$, $p(0)=\ed\ms{V}^\text{eq}[m_0^*]/\ed m$ as initial conditions to generate the characteristic $[q(s),p(s)]_{0\leq s\leq t}$ with error $\eps=|q(t)-m|$. 

Iterating this procedure for an ever finer grid on an ever smaller interval around the exact value of $m_0$, we can achieve an error smaller than a given threshold, $\eps_\text{min}$. We use the standard ODE solver ``ode45'' in MatLab with non-default accuracy settings 'RelTol'=$10^{-12}$ and 'AbsTol'=$10^{-16}$ to integrate \Eqsref{bvp}. We take $M=10^2$ and devise an iterative procedure that reduces the interval size around $m_0$ by a factor of five in each iteration, until an error smaller than $\eps_\text{min}=10^{-10}$ is achieved.

This program is repeated for a grid of $m$ and $t$ values for each of which $V(m,t)$ is computed from Eq.~(9) in the main text. The two other fields, $\partial_mV(m,t)$ and $m_0(m,t)$, are evaluated directly from the end and starting points of the characteristics, $p(t)$ and $q(0)$, respectively.
\section{Derivation and solution of Riccati equation}\seclab{ric}
Here we derive an equation for the curvature $z(t)=\partial_m^2V(0,t)$ at $m=0$, which tends to negative infinity at time $t_\text{c}$ when $V(m,t)$ develops a kink at $m=0$. To obtain this equation, we use $p(s) = \partial_m V[q(s),s]$ to write the Hamilton equation $\dot p = -\partial_q\ms{H}(q,p)$ as
\algn{
	\partial^2_m V(m,t)\partial_p\ms{H}[m,\partial_mV(m,t)] + \partial_m\partial_t V(m,t) = -\partial_q\ms{H}[m,\partial_mV(m,t)]\,,
}
where we evaluated the equation at $t$ with $q(t)=m$. We take a partial derivative with respect to $m$ to obtain
\begin{multline}\eqnlab{longveqn}
	\partial^3_m V(m,t)\partial_p\ms{H}[m,\partial_mV(m,t)] + 	\partial^2_m V(m,t)\{\partial_q\partial_p\ms{H}[m,\partial_mV(m,t)] +\partial^2_p\ms{H}[m,\partial_mV(m,t)]\partial^2_mV(m,t)\} + \partial^2_m\partial_t V(m,t) =\\ -\partial^2_q\ms{H}[m,\partial_mV(m,t)]-\partial_p\partial_q\ms{H}[m,\partial_mV(m,t)]\partial^2_mV(m,t)\,,
\end{multline}
We swap the order of the partial derivatives and evaluate \eqnref{longveqn} along a trajectory $m=q(s)$ at $t=s$ to obtain an equation for $Z(s)=\partial_m^2V[q(s),s]$. We use
\algn{
	\dot Z(s)=\tfrac{\ed}{\ed s}\partial_m^2V[q(s),s] = \partial^3_m V[q(s),t]\dot q(s) + \partial_s \partial^2_m V[q(s),s]\,,
}
to write \Eqnref{longveqn} as the Riccati equation
\algn{\eqnlab{ricgen}
	\dot Z(s) = -\partial_q^2\ms{H}[q(s),p(s)]-2\partial_q\partial_p\ms{H}[q(s),p(s)]Z(s)-\partial_p^2\ms{H}[q(s),p(s)]Z(s)^2\,,
}
with initial condition $Z(0) = \partial_m^2V[q(0),0]=\tfrac{\ed^2}{\ed m^2}\ms{V}^\text{eq}[q(0)]$. Along the trajectory $q(s)=p(s)=0$ and evaluating at $s=t$ this simplifies to an equation for $z(t)=\partial_m^2V(0,t)$,
\algn{\eqnlab{ric}
	\dot z(t) = -\partial_q^2\ms{H}(0,0)-2\partial_q\partial_p\ms{H}(0,0)z(t)-\partial_p^2\ms{H}(0,0)z(t)^2\,.
}
Using $\partial_q^2\ms{H}(0,0)=0$, $\partial_q\partial_p\ms{H}(0,0) = 2(\beta_\text{q}J-1)/\tau=-2J(\beta_\text{c}-\beta_\text{q})/\tau=$ and $\partial_p^2\ms{H}(0,0)=4/\tau$, we obtain the Riccati equation referred to in the main text, with initial condition
\algn{
	z(0)=\partial_m^2V(0,0)=\tfrac{\ed^2}{\ed m^2}\ms{V}^\text{eq}(0)=-J(\beta-\beta_\text{c})\,.
}
The solution $z(t)$ to \Eqnref{ric} then reads
\algn{\eqnlab{ricsol}
	z(t)= \frac{J(\beta-\beta_\text{c}) (\beta_\text{c}-\beta_\text{q})}{\beta-\beta_\text{c}-(\beta-\beta_\text{q}) e^{-4J (\beta_\text{c}-\beta_\text{q}) t}}\,.
}
When $\beta>\beta_\text{c}$ and $\beta_\text{q}<\beta_\text{c}$, i.e., a disordering quench, the denominator in \Eqnref{ricsol} tends to zero, and $z(t)$ tends to $-\infty$, at the finite time 
\algn{\eqnlab{tcCW}
	t_\text{c} = \frac{\tau}{4J(\beta_\text{c}-\beta_\text{q})}\ln \left[\frac{\beta-\beta_\text{q}}{\beta-\beta_\text{c}}\right]\,,
}
as stated in the main text. Unstable gradients descibed by Riccati equations are common in problems where so-called ``caustics'' form~\cite{Kul82,Wil05,Mei20}. Conventional caustics are singularities caused by the partial focusing of light in ray optics~\cite{Ber80}. In the present problem, analogous caustics occur as partial focuses of the characteristics $[q(s),p(s)]_{0\leq s\leq t}$ for varying $q(0)=m_0$, whenever characteristics cross in space, indicated by $Z(s)\to-\infty$. Because points of crossing characteristics correspond to the phase boundaries in our problem, we use \Eqnref{ricgen} together with the Hamilton equations to obtain numerically accurate dynamical phase diagrams, Figs.~3(a) and 3(f) in the main text.

The instability of the Riccati equation \eqnref{ric} for $z(t)$ can be visualised by writing $\dot z(t) = -W'[z(t)]$ with potential
$W(z) = 4/(3\tau) z^3 - 2J(\beta_\text{c}-\beta_\text{q})/\tau z^2$. Figure~\figref{fig1} shows $W(z)$, with stable and unstable fixed points shown as the red and green dots, respectively. Since $W(z)$ is decreasing for $z\leq0$, any negative initial $z(0)$ leads to $z(t)\to-\infty$, at the finite time $t_\text{c}$, and $V(m,t)$ forms a kink at zero. For any positive initial $z(0)$, on the other hand, $z(t)$ tends to the stable fixed point (green bullet), corresponding to the curvature of $\ms{V}_\text{q}^\text{eq}(m)$ at $m=0$.
\begin{figure}
	\includegraphics[]{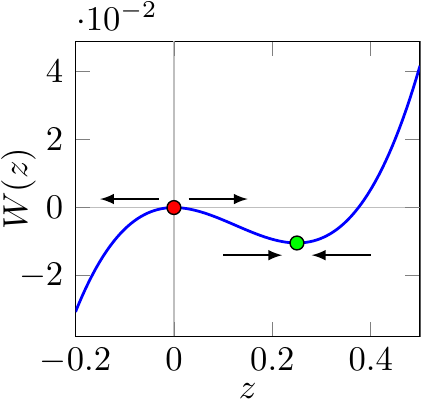}
	\caption{Potential $W(z)$ for $\beta=5/(4J)$ and $\beta_\text{q}=3/(4J)$. Arrows indicate the (in)stability of the fixed points (coloured bullets).}\figlab{fig1}
\end{figure}
For a disordering quench with $\beta_\text{q}<\beta_\text{c}<\beta$, $z(0)$ is always negative, so that the kink of $V(m,t)$ at $m=0$ always forms.
\section{Dynamical Landau potential}\seclab{dlp}
In this section, we give a more detailed derivation of the dynamical Landau potential $\ms{L}^{m,t}(m_0)$ for scalar, parity-symmetric order parameters. As the observable, we consider a parameter $m$, whose rate function $V(m,t)$ is parity symmetric $V(m,t) = V(-m,t)$ and obeys the Hamilton-Jacobi equation
\algn{\eqnlab{hjeqngen}
	\partial_tV(m,t) + \ms{H}[m,\partial_mV(m,t)]=0\,,\qquad\text{with initial condition}\qquad V(m,0) \sim \frac{\xi_0}2 m^2 + \frac{\xi_1}4 m^4-V_\text{min}\,,
}
where $V_\text{min}$ is a constant that ensures $\min_{m}V(m,0)=0$. For the Curie-Weiss model discussed in the main text, $V_\text{min} = \beta\ms{F}(\bar m)$. We assume $\xi_0<0$ and $\xi_1>0$ so that the initial state is ordered. In the long-time limit, the system relaxes to a disordered state and the asymptotic rate function $\lim_{t\to\infty}V(m,t)$ has a single minimum at $m=0$. The Hamilton-Jacobi equation~\eqnref{hjeqngen} with Hamiltonian $\ms{H}(q,p)$, is solved by the Hamilton equations $\dot q = \partial_p\ms{H}(q,p)$, $\dot p = -\partial_q\ms{H}(q,p)$ with the boundary conditions $p(0) = \partial_mV[q(0),0] \sim \xi_0 q(0) + \xi_1 q(0)^3$ and $q(t) = m$.

As we show in detail below, the dynamical Landau potential $\ms{L}^{m,t}(m_0)$ fulfils, close to the critical point, the relation
\algn{\eqnlab{Vpot}
	V(m,t) =\min_{m_0}\ms{L}^{m,t}(m_0) -V_\text{min}\,,\qquad\text{with}\qquad\ms{L}^{m,t}(m_0) \sim - m\lambda_0 m_0 -\lambda_1\left(\frac{t-t_\text{c}}\tau\right)\frac{m_0^2}2  + \left[\lambda_2+\lambda_3\left(\frac{t-t_\text{c}}\tau\right)\right]\frac{m_0^4}4\,.
}
For $m=0$, and at the critical time $t_\text{c}$, $\ms{L}^{m,t}(m_0)$ transitions from a single into a double well when $\lambda_1(t_\text{c})=0$. A dynamical equation of state
\algn{\eqnlab{deos}
	m \sim -\left[\frac{\lambda_1}{\lambda_0}\left(\frac{t-t_\text{c}}\tau\right)+\ldots\right]m_0 + \left(\frac{\lambda_2}{\lambda_0}+\ldots\right)m_0^3\,,
}
for $|t-t_\text{c}|\ll\tau$ follows from the minimisation condition $\partial_{m_0}\ms{L}^{m,t}(m_0)=0$.

In order to derive the asymptotic expression in \Eqsref{Vpot}, we start with a general fourth-order Hamiltonian $\ms{H}$ that is invariant under the parity operation $(q,p)\to(-q,-p)$:
\algn{\eqnlab{hamun}
	\ms{H}(q,p) = \ms{H}_0(q,p) + \ms{H}_1(q,p)\,,
}
where
\algn{\eqnlab{hamun1}
	\ms{H}_0(q,p) = \frac12 a_1 p^2 + a_2 qp\,\qquad\text{and}\qquad \ms{H}_1(q,p) = b_1 p^4 + b_2 p^3 q + b_3 p^2 q^2 + b_4 p q^3\,.
}
Note that the terms $\propto p^2$ in $\ms{H}_0$ and $\propto p^4$ in $\ms{H}_1(q,p)$ must vanish to guarantee the existence of a deterministic noiseless limit, for which $\dot p|_{p=0} = -\partial_p\ms{H}(q,0) = 0$. We also require $a_1>0$ and $a_2<0$, to ensure that in the noiseless limit, $p=0$ is the stable manifold of the fixed point $(q,p)=(0,0)$, and that the asymptotic rate function $\lim_{t\to\infty}V(m,t)$ has a single minimum at $m=0$.

We first show that any Hamiltonian $\ms{H}$ of the form given in \Eqsref{hamun} and \eqnref{hamun1} can be brought into the form
\algn{\eqnlab{ham}
	\ms{H}'(Q,P) \sim \frac12 \left(P^2 - Q^2\right) + \frac{\alpha}4\left(	P^4 + Q^4	\right)\,,
}
stated in the main text, with real parameter $\alpha$. This is done by means of canonical transformations and a time rescaling, neglecting higher-order terms. To this end, we first remove the mixed term ($\propto qp$) in $\ms{H}_0(q,p)$ by the linear canonical transformation
\algn{
	\pmat{q'\\p'}= \pmat{\omega_0&0	\\-\omega_0&\omega_0^{-1}}\pmat{q\\p}\,,
}
where $\omega_0 = \sqrt{|a_2/a_1|}>0$. This gives $\ms{H}_0(q',p') = |a_2|(p'^2 - q'^2)/2$ where $\tau_a = 1/| a_2|$ is a time scale. We dedimensionalise time according to $t\to t' = t/\tau_{a}$ leading to the Hamilton-Jacobi equation $0=\partial_{t'} V(q',t') + \tau_a\ms{H}(q',\partial_{q'} V)$ with the dimensionless quadratic and quartic Hamiltonians
\algn{\eqnlab{hprime}
	\ms{H}'_0(q',p') = \tau_a \ms{H}_0(q',p') = \frac12\left(p'^2 - q'^2\right)\,,\qquad \ms{H}'_1(q',p') = \tau_a \ms{H}_1(q',p') = b'_1 p'^4 + b'_2 p'^3 q' + b'_3 p^2 q^2 + b'_4 p q^3 + b'_5 q^4\,,
}
respectively, and with the coefficients
\algn{
	b'_1 =& -\frac{a_2 b_1}{a_1^2}\,,\quad b'_2 = \frac{a_1 b_2-4 a_2 b_1}{a_1^2}\,,\quad b'_3 = \frac{3 (a_1 b_2-2 a_2 b_1)}{a_1^2}-\frac{b_3}{a_2}\,,\\
	\quad b'_4 =& -\frac{4 a_2 b_1}{a_1^2}+\frac{3 b_2}{a_1}+\frac{a_1 b_4}{a_2^2}-\frac{2 b_3}{a_2}\,,\quad b'_5=-\frac{a_2 b_1}{a_1^2}+\frac{b_2}{a_1}+\frac{a_1 b_4}{a_2^2}-\frac{b_3}{a_2}\,.
}
Note that $\ms{H}'_0$ is now in the desired form given in \Eqnref{ham}. In the next step, we perform a non-linear canonical transformation to simplify the quartic Hamiltonian $\ms{H}'_1(q',p')$. We use the transformations
\algn{\eqnlab{canQP}
	Q \sim q' + \Psi^{(3)}(q',p')\qquad\text{and}\qquad P \sim p' + \Phi^{(3)}(q',p')\,,
}
where $\Psi^{(3)}$ and $\Phi^{(3)}$ are third order in $q'$ and $p'$; higher orders are neglected. To this order, the inverse transformations are given by
\algn{\eqnlab{canqp}
	q' \sim Q - \Psi^{(3)}(Q,P)\qquad\text{and}\qquad p' \sim P - \Phi^{(3)}(Q,P)\,.
}
The functions $\Psi^{(3)}$ and $\Phi^{(3)}$ have the general form
\algn{
	\Phi^{(3)}(q',p') = c_1 p'^3 + c_2 q p'^2 + c_3 q'^2 p' + c_4 q'^3\,, \qquad 	\Psi^{(3)}(q',p') = c_0 p'^3 -3 c_1 q' p'^2 - c_2 q'^2 p' -\frac13 c_3 q'^3\,.
}
This choice of $\Phi^{(3)}$ and $\Psi^{(3)}$ defines a canonical transformation for arbitrary parameters $c_0,\ldots,c_4$. The transformation depends only on these five parameters, due to the requirement that a canonical transformation must leave the Hamilton equations invariant~\cite{Gol80}. Note also that \Eqsref{canQP} are trivial to first order in $q'$ and $p'$ so that the quadratic Hamiltonian $\ms{H}'_0$ is left invariant, thus retaining its simple form. Substituting \eqnref{canqp} into \Eqnref{hprime} and demanding that $\ms{H}'_1(Q,P)$ be of the form specified in \Eqnref{ham} we determine coefficients $c_1,\ldots,c_4$:
\algn{
	c_1= \frac{1}{6} (3 b'_1+b'_3-3 b'_5)\,,\quad c_2= b'_2+c_0\,,\quad c_3= \frac{1}{2} (-3 b'_1+b'_3+3 b'_5)\,,\quad c_4=-b'_2+b'_4-c_0\,,
}
or
\algn{
	c_1= \frac{-6 a_2^3 b_1+2 a_1^2 a_2 b_3-3 a_1^3 b_4}{6 a_1^2 a_2^2}\,,\quad c_2= \frac{a_1 b_2-4 a_2 b_1}{a_1^2}+c_0\,,\quad c_3= -\frac{3 a_2 b_1}{a_1^2}+\frac{3 b_2}{a_1}+\frac{3 a_1 b_4}{2 a_2^2}-\frac{2 b_3}{a_2}\,,\quad c_4=\frac{2 b_2}{a_1}+\frac{a_1 b_4}{a_2^2}-\frac{2 b_3}{a_2}-c_0\,.
}
The expression for $\alpha$ in \Eqnref{ham} then reads
\algn{
	\alpha =2\left(b'_1-\frac13 b'_3+b'_5\right)=\frac{2 (3 a_1 b_4-2 a_2 b_3)}{3 a_2^2}\,.}
The parameter $c_0$ can be chosen freely. For convenience, we set $c_0=0$ which ensures that $q$ vanishes whenever $Q=0$, and vice-versa.

Canonical transformations leave the Hamilton equations invariant, so that the same equations hold for the transformed coordinates $(Q,P)$. However, we must still transform the initial conditions $q(0)=m_0$ and $p(0) = \xi_0 m_0 + \xi_1 m_0^3$, which leads to
\algn{
	Q(0)\sim\omega_0m_0 + \omega_1m_0^3\qquad\text{and}\qquad P(0)\sim \sigma_0 m_0 + \sigma_1 m_0^3 \,, 
}
where
\algn{
	\sigma_0 =& \frac{\xi_0}{\omega_0}-\omega_0\,,\qquad \sigma_1 = \frac{c_1 \left(\xi_0-\omega_0^2\right)^3+\omega_0^2 \left[c_3 \xi_0 \omega_0^2+c_2 \left(\xi_0-\omega_0^2\right)^2-c_3 \omega_0^4+c_4 \omega_0^4+\xi_1\right]}{\omega_0^3}\,,\\
	\omega_1 =& -\frac{3 c_2 \xi_0 \omega_0^4-3 c_0 \left(\xi_0-\omega_0^2\right)^3+9 c_1 \left(\omega_0^3-\xi_0 \omega_0\right)^2-3 c_2 \omega_0^6+c_3 \omega_0^6}{3 \omega_0^3}\,.
}
From these equations, note that since $\xi_0<0$ and $\omega_0>0$ we have $\sigma_0<0$ and $\sigma_0+\omega_0<0$. This derivation shows that an arbitrary dynamical Hamiltonian of the form \eqnref{hamun}--\eqnref{hamun1} can be brought into the much simpler form \eqnref{ham} by canonical transformations and a time rescaling.

We now establish the connection between $V(m,t)$ and $\ms{L}^{m,t}(m_0)$, as well as the asymptotic expression for $\ms{L}^{m,t}(m_0)$ in \Eqsref{Vpot}. Due to the variational principle for $V(m,t)$, $\delta V(m,t)=0$, we can write $V(m,t)$ as minimisation over trajectories $v(s)$ and $w(s)$:
\algn{\eqnlab{vvar}
	V(m,t) = \min_{\substack{v(s),w(s)\\v(t)=m}}\left\{\int_0^t \ed s \left[ w\dot v  - \ms{H}(v,w) \right] + \ms{V}^\text{eq}[v(0)]	\right\} = \min_{m_0}\min_{\substack{v(s),w(s)\\v(0)=m_0,v(t)=m}}\left\{\int_0^t \ed s \left[ w \dot v  - \ms{H}(v,w) \right] + \ms{V}^\text{eq}[v(0)]	\right\}\,.
}
We now write the inner minimum in \Eqsref{vvar} as
\algn{\eqnlab{lvar}
\ms{L}^{m,t}(m_0)-V_\text{min}\equiv\min_{\substack{v(s),w(s)\\v(0)=m_0,v(t)=m}}\left\{\int_0^t \ed s \left[ w \dot v  - \ms{H}(v,w) \right] + \ms{V}^\text{eq}[v(0)]	\right\}\,,
}
which gives $V(m,t) =\min_{m_0}\ms{L}^{m,t}(m_0) -V_\text{min}$, as stated in \Eqsref{Vpot}. The characteristics $[q(s),p(s)]_{0\leq s\leq t}$ that minimise the expression in \Eqnref{lvar} are solutions to the same Hamilton equations, but with boundary conditions $q(0)=m_0$ and $q(t)=m$. In terms of these characteristics, we express $\ms{L}^{m,t}(m_0)$ as
\algn{\eqnlab{landau}
	\ms{L}^{m,t}(m_0) = \int_{m_0}^m p(q)\ed q - t\ms{H}[m_0,\partial_qV(m_0,0)] + V(m_0,0) + V_\text{min}\,,
}
where we have used that the Hamiltonian is constant during the time evolution, and that we can express $p$ in terms of $q$ using the Hamiltonian $\ms{H}$. The term involving $\ms{H}$ in \Eqnref{landau} is straightforwardly calculated, because it only depends on the initial conditions. We find
\algn{
	 t\ms{H}[m_0,\partial_qV(m_0,0)] \sim& t'\ms{H}'(\omega_0m_0+\omega_1m_0^3,\sigma_0m_0 + \sigma_1m_0^3)\,,\\\
	\sim& \frac{t'}{2} m_0^2 \left(\sigma_0^2-\omega_0^2\right)-t' m_0^4 \left[\omega_0 \omega_1-\sigma_0 \sigma_1-\frac{1}{4} \alpha  \left(\sigma_0^4+\omega_0^4\right)\right]\,,
}
and $V(m_0,0)$ given in \Eqsref{hjeqngen}. We are now left with evaluating the first term in \Eqnref{landau} up to fourth order in $m_0$ and to lowest order in $m$. Expressing the argument of the integral in terms of $Q$ and $P$ and carefully keeping track of the orders in $m_0$ we finally obtain
\begin{multline}\eqnlab{Lpot}
	\ms{L}^{m,t}(m_0) \sim -m  \omega_0 \sqrt{\sigma_0^2-\omega_0^2}m_0- \left(\sigma_0^2-\omega_0^2\right) \left(\frac{t-t_\text{c}}{\tau_a}\right)\frac{m_0^2}{2}  \\ +\left\{\omega_0  \sigma_1-\omega_1\sigma_0 -\frac{\alpha}{4}
 \left[3  \left(\sigma_0^2-\omega_0^2\right)^2\frac{t_\text{c}}{\tau_a}-\sigma_0 \omega_0 \left(\sigma_0^2+\omega_0^2\right)\right]+ \left[4(\omega_0 \omega_1-\sigma_0 \sigma_1)-\alpha  \left(\sigma_0^4+\omega_0^4\right)\right]\left(\frac{t-t_\text{c}}{\tau_a}\right)\right\}\frac{m_0^4}4 + \ms{O}(m_0^6)\,,
\end{multline}
which is of the form stated in \Eqsref{Vpot}, allowing us to read off the coefficients $\lambda_{0,1,2,3}$. The critical time $t_c$ where $\ms{L}^{m,t}(m_0)$ transitions from a single into a double well potential for $m=0$ is given by
\algn{\eqnlab{tceqn}
	t_\text{c} = \frac{\tau_a}{2}\ln\left[\frac{\sigma_0-\omega_0}{\sigma_0+\omega_0}\right]\,.
}
Remarkably, the dynamical Landau potential $\ms{L}^{m,t}(m_0)$ depends solely on the five parameters $\alpha$, $\omega_0$, $\omega_1$, $\sigma_0$ and $\sigma_1$, although it applies to the general, fourth-order Hamiltonian $\ms{H}$ given in \Eqsref{hamun} and \eqnref{hamun1}. The dynamical equation of state follows by expressing the minimisation condition $\partial_{m_0}\ms{L}^{m,t}(m_0)=0$ in terms of $m$.

From the general form of the dynamical Landau potential in \Eqnref{Lpot}, and the corresponding critical time $t_\text{c}$ in \Eqnref{tceqn}, we observe that $\sigma_0<0$, $\omega_0>0$, and $\sigma_0+\omega_0<0$ are sufficient to ensure the occurrence of the dynamical phase transition at time $t_\text{c}$. These conditions hold whenever the order parameter is initially in an ordered state, $\xi_0<0$, and quenched into a disordered state, requiring $a_1>0$ and $a_2<0$.

We now calculate the coefficients for the dynamical Landau potential $\ms{L}^{m,t}(m_0)$ in \Eqnref{Lpot} for the Curie-Weiss model and for a Brownian particle subject to a potential quench in the weak-noise limit.
\subsection{Curie-Weiss model}
We first compute the five parameters $\alpha$, $\omega_0$, $\omega_1$, $\sigma_0$ and $\sigma_1$ that determine the dynamical Landau potential $\ms{L}_\text{CW}^{m,t}(m_0)$ for the Curie-Weiss model. Expanding the Hamiltonian
\algn{\eqnlab{Hcw}
	\ms{H}(q,p) = w_+(q) (\ee^{2p}-1) + w_-(q) (\ee^{-2p}-1)\,, 
} 
Eq.~(6) in the main text, to fourth order in $q$ and $p$, we obtain the coefficients $a_1$, $a_2$ and $b_1,\ldots,b_5$. A fourth order expansion of the equilibrium rate function $\ms{V}_\text{eq}$ at inverse temperature $\beta$ gives $\xi_0$ and $\xi_1$. From these coefficients, we obtain the parameters
\sbeqs{\eqnlab{params}
\algn{
	\alpha =& -\frac{4 \beta_\text{q} }{3 J(\beta_\text{c}-\beta_\text{q})^2}\,,\quad \omega_0 = \sqrt{\frac{J(\beta_\text{c}-\beta_\text{q})}2}\,,\quad \sigma_0 = -\frac{2J\left(\beta-\frac{\beta_\text{q} +\beta_\text{c}}2\right)}{\sqrt{2J(\beta_\text{c}- \beta_\text{q})}}\\
	\omega_1 =&\frac{J}{3 [2J(\beta_\text{c}- \beta_\text{q})]^{5/2}}\bigg[6 \beta ^2 J \left(\beta_\text{q}^3 J^3-3 \beta_\text{q}^2 J^2-\beta_\text{q} J-1\right)+6 \beta  \left(-\beta_\text{q}^4 J^4+2 \beta_\text{q}^3 J^3+4 \beta_\text{q}^2 J^2+2 \beta_\text{q} J+1\right)\nn\\
	&+\beta_\text{q} \left(\beta_\text{q}^4 J^4+\beta_\text{q}^3 J^3-13 \beta_\text{q}^2 J^2-5 \beta_\text{q} J-8\right)\bigg]\,,\\
	\sigma_1 =& \frac{2 J}{3 [2J(\beta_\text{c}- \beta_\text{q})]^{7/2}} \bigg[4 \beta ^3 J^2 \left(\beta_\text{q}^3 J^3-3 \beta_\text{q}^2 J^2-\beta_\text{q} J-1\right)+6 \beta ^2 J \left(-\beta_\text{q}^4 J^4+2 \beta_\text{q}^3 J^3+4 \beta_\text{q}^2 J^2+2 \beta_\text{q} J+1\right)\nn\\
	&+6 \beta  \left(2 \beta_\text{q}^4 J^4-7 \beta_\text{q}^3 J^3+\beta_\text{q}^2 J^2-5 \beta_\text{q} J+1\right)+\beta_\text{q} \left(\beta_\text{q}^5 J^5-6 \beta_\text{q}^4 J^4+8 \beta_\text{q}^3 J^3+2 \beta_\text{q}^2 J^2+15 \beta_\text{q} J-4\right)\bigg]\,,
}
}
and the time scale $\tau_a = \tau/[2J(\beta_\text{c}-\beta_\text{q})]$. For the values $\beta=5/(4J)$ and $\beta_\text{q}=3/(4J)$ used to create Figs.~2 and 3 in the main text, the parameters take the values
\algn{
	\alpha = -16\,,\quad \omega_0 = \frac{1}{2 \sqrt{2}}\,,\quad \omega_1=-\frac{831}{128 \sqrt{2}}\,,\quad \sigma_0 = -\frac{3}{2 \sqrt{2}}\,,\quad\sigma_1=-\frac{1337}{384 \sqrt{2}}\,,\quad\tau_a = 2\tau\,.
}
With help of the coefficients in \Eqnref{params}, we obtain the dynamical Landau potential for the Curie-Weiss model as
\algn{\eqnlab{dlpCW}
	\ms{L}^{m,t}_\text{CW}(m_0) = -m J\sqrt{(\beta -\beta_\text{q}) (\beta-\beta_\text{c})}m_0 - 4 J^2 (\beta -\beta_\text{q}) (\beta  -\beta_\text{c})\left(\frac{t-t_\text{c}}\tau\right)\frac{m_0^2}2 + \left[\lambda_2 + \lambda_3\left(\frac{t-t_\text{c}}\tau\right)	\right]\frac{m_0^4}4\,,
}
with
\algn{
	\lambda_2 =& \frac{\beta -\beta_\text{q}}{3 J(\beta_\text{c}-\beta_\text{q})^3} \bigg\{\beta_\text{q}^4 J^3 (1-\beta  J)+2 \beta_\text{q}^3 J^2 \left(\beta ^2 J^2-1\right)+\beta  (3-2 \beta  J)-\beta_\text{q}^2 J \left[3 \beta  J (2 \beta  J-3)+24 t_\text{c} (\beta  J-1)^2+2\right]\nn\\
	&+\beta_\text{q} \left[24 \beta  J   (\beta  J-1)^2 t_\text{c}+\beta  J-3\right]\bigg\}\,,\\
	\lambda_3 =& -\frac{4J^3(\beta -\beta_\text{q})}{3}\left[-\beta_\text{q}^2+2 \beta ^3 J-2 \beta ^2 (\beta_\text{q} J+2)+\beta  \beta_\text{q} (\beta_\text{q} J+2)\right]\,,
}
and $t_\text{c}$ given in \Eqnref{tcCW}. For the values $\beta=5/(4J)$ and $\beta_\text{q}=3/(4J)$ we obtain for the parameters $\lambda_0,\ldots,\lambda_3$:
\algn{\eqnlab{lamvals}
	\lambda_0 = \frac{1}{2 \sqrt{2}}\,,\qquad \lambda_1 = \frac12\,,\qquad\lambda_2=6\ln(2)-\frac{371}{96}\,,\qquad\lambda_3 = \frac{57}{32}\,,
}
and for the critical time $t_\text{c}/\tau=\ln(2)\approx 0.693$, as stated in the main text. Figure~\figref{fig2}(a) shows how $\ms{L}^{m,t}(m_0)$ transitions from a single into a double-well at the critical time $t_\text{c}$, for the parameter values in \Eqsref{lamvals} and $m=0$. The dynamical order parameter $m_0(m,t)$ is given by the minima of $\ms{L}^{m,t}(m_0)$.

\begin{figure}
	\includegraphics{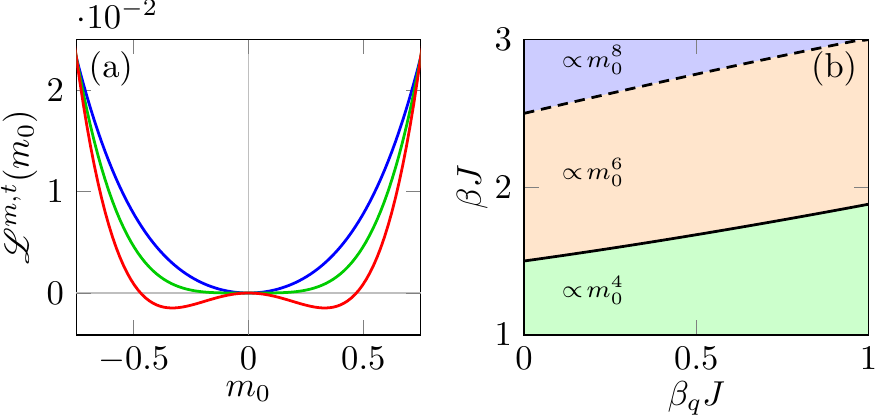}
	 \caption{(a) Dynamical Landau potential $\ms{L}^{m,t}(m_0)$ for the Curie-Weiss model with $m=0$, and parameters $\beta=5/(4J)$ and $\beta_\text{q} =3/(4J)$, evaluated at $t/\tau=0.6$ (blue), $t/\tau=\ln(2)$ (green), and $t/\tau=0.8$ (red). (b) Values for $\beta_\text{q}$ and $\beta$ where terms of order $m_0^4$, $m_0^6$ and $m_0^8$ must be included in $\ms{L}^{m,t}_\text{CW}(m_0)$, \Eqnref{dlpCW}.}\figlab{fig2}
\end{figure}

Both $\lambda_1$ and $\lambda_0$ are positive for $\beta<3/(2J)$. We note, however, that for $\beta>3/(2J)$, $\lambda_2$ is negative for some values of $\beta_\text{q}<\beta_\text{c}$, and becomes entirely negative for $\beta\gtrapprox1.884$. When $\lambda_2$ is negative, the next-higher order $\propto C m_0^6$ in $\ms{L}^{m,t}(m_0)$ needs to be taken into account. It turns out that $C>0$ for $\beta<5/(2J)$, but similarly to $\lambda_2$, $C$ becomes negative for $\beta>5/(2J)$, so that terms of order $m_0^8$ must be considered. The regions in $(\beta_\text{q},\beta)$-space where different powers of $m_0$ must be included in $\ms{L}^{m,t}(m_0)$ are shown in \Figref{fig2}(b). At the boundary between the $\propto m_0^4$ and $\propto m_0^6$ regions (solid line) the dynamical phase transition changes from continuous to first order.
\subsection{Potential quench for Brownian particle at weak-noise}
We now show the occurrence of the finite-time dynamical phase transition for a Brownian particle in the weak-noise limit, that experiences a quench of the potential. Consider an overdamped particle at position $x$, subject to an external potential $\ms{W}(x)$, and immersed in a heat bath at inverse temperature $\beta=1/(\kb T)$. For $t<0$, $\ms{W}(x)$ is a symmetric double-well potential $\ms{W}_\text{eq}(x)=\lambda_0(x^2-x_\text{m}^2)^2/4$ with coupling strength $\lambda_0$ and potential minima at $\pm x_\text{m}$. The potential is quenched from a double into a single-well, $\ms{W}_\text{eq}\to\ms{W}_\text{q}=\lambda_\text{q}x^2/2$, at time $t=0$; $\lambda_\text{q}$ denotes the harmonic coupling. After the quench, the system relaxes into a new equilibrium state, determined by $\ms{W}_\text{q}$.

The weak-noise limit is attained when the thermal energy $\kb T$ is much smaller than both $\ms{W}_\text{eq}(0)$ and $\ms{W}_\text{q}(x_\text{m})$, while the ratio $\Gamma \equiv \ms{W}_\text{eq}(0)/\ms{W}_\text{q}(x_\text{m})$ remains finite. We dedimensionalise the spatial coordinate $x\to x_\text{m} x$ and find that in the weak-noise limit, the initial equilibrium probability density has the initial rate function $V(x,0) = \Gamma (x^2-1)^2$. The inverse weak-noise parameter $\eps^{-1}=\ms{W}_\text{q}(x_\text{m})/(\kb T)\gg1$ takes the role of $N$ in the thermodynamic limit. Due to the double-well shape of $\ms{W}_\text{eq}$, and thus of $V(x,0)$, the system is initially in an ordered state. The dynamical order parameter is given by the (dimensionless) initial position $q(0) = x_0(x,t)$ of the optimal fluctuation $q(s)$, $0\leq s \leq t$, that achieves the particle position $q(t) = x$ at final time $t$. By taking the weak-noise limit $\eps\ll1$ of the associated Fokker-Planck equation, we find that the post-quench dynamics of the rate function $V(x,t)$ is determined by the Hamiltonian-Jacobi equation
\algn{
	0=\partial_tV(x,t) +\ms{H}_\text{FP}[x,\partial_xV(x,t)]\,, \qquad \ms{H}_\text{FP}(q,p) = \frac1{\tau_\text{d}}\left(p^2 - pq\right)\,,
}
where $\tau_\text{d}=(\lambda_\text{q} \mu)^{-1}$ is the damping time scale, $\mu$ denotes the mobility. From the model parameters, we obtain the coefficients $\tau_a = \tau_\text{d}$, $\alpha = \omega_1= 0$, and
\algn{
	\omega_0 = \frac1{\sqrt{2}}\,,\qquad	\sigma_1 = -\frac{8 \Gamma +1}{\sqrt{2}}\qquad \sigma_2 = 4 \sqrt{2} \Gamma\,.
}
Using \Eqnref{Lpot}, this immediately leads us to the dynamical Landau potential
\algn{\eqnlab{lwn}
	\ms{L}_\text{WN}^{x,t}(x_0) \sim -2 x \sqrt{\Gamma  (4 \Gamma +1)}x_0  -8 \Gamma  (4 \Gamma +1)\left(\frac{ t-t_\text{c} }{\tau_\text{d}}\right)\frac{x_0^2}2 + \left[4\Gamma +16 (8 \Gamma +1) \Gamma\left(\frac{ t-t_\text{c} }{\tau_\text{d}}\right)\right]\frac{x_0^4}4\,,
}
with critical time $t_\text{c}/\tau = \log \left(\frac{1}{4 \Gamma }+1\right)/2$. Because $\lambda_{0,1,2}>0$ and $t_\text{c}>0$ in \Eqnref{lwn} [c.f. \Eqnref{Vpot}], this shows that the finite-time dynamical phase transition occurs also in this setting.
\section{Robustness of dynamical phase transition}
In this section, we show that the dynamical phase transition is robust against transformations of the microscopic transition rates $W_{\pm}(M)$, given in \Eqsref{mastereqn} [Eq.~(3) in the main text], that preserve detailed balance and the parity symmetry of the problem:
\algn{\eqnlab{conds}
	W_\pm(M)P^\text{eq}(M)=W_\mp(M_\pm)P^\text{eq}(M_\pm)\quad \text{and} \quad W_{\pm}(M)|_{H=0}=W_{\mp}(-M)|_{H=0}\,.
}
These requirements do not determine $W_{\pm}(M)$ uniquely. For any (local) transformation of the rates 
\algn{
W_\pm(M)\to\tilde W_{\pm}(M)=W_{\pm}(M) F_{\pm}(M)\,,
}
with given functions $F_{\pm}$, the same conditions \eqnref{conds} are satisfied by $\tilde W_\pm(M)$ as long as 
\algn{\eqnlab{fprops}
F_\pm(M) = F_\mp(M_\pm)=F_\mp(M\pm2)\quad\text{and}\quad F_\pm(M)=F_\mp(-M)\,.
}
As an example, consider the transformation
\algn{\eqnlab{ftrafo}
	F_\pm(M) = \frac{\ee^{\pm\beta[J(M\pm1)+H]}}{1+\ee^{\pm2\beta[J(M\pm1)+H]}}\,,
}
that brings Arrhenius-type rates, as given in \Eqsref{mastereqn} [Eq.~(3) in the main text], into the Glauber-type form
\algn{\eqnlab{wtilde}
	\tilde W_{\pm}(M) = \frac{N\mp M}{2\tau}\frac{\ee^{\pm2\beta[J(M\pm1)+H]}}{1+\ee^{\pm2\beta[J(M\pm1)+H]}}\,.
}
The expression in \Eqnref{ftrafo} satisfies the requirements in \Eqsref{fprops}, so that the transformed rates \eqnref{wtilde} obey detailed balance and are parity symmetric. We now show that the occurrence of the dynamical phase transition is robust against any such transformation $F_\pm(M)$.

To this end, we define $f_\pm(m)\equiv F_\pm(Nm)$ and note that as $N\to\infty$ one has $f_\pm(m)= F_\pm(Nm)=F_\mp[N(m\pm2/N)]\sim F_{\mp}(Nm)=f_{\mp}(m)$, so that $f_\pm(m) = f_\mp(m)\equiv f(m)$. In other words, for $N\to\infty$, $f$ must be independent of the direction $\pm$ of the transition.

Second, the symmetry condition $F_\pm(M)=F_\mp(-M)$ shows that $f$ is even, $f(m) = f(-m)$, for $H=0$. Third, a constant $f(m)=c$ leads to a mere rescaling of the microscopic transition time $\tau$, so that we may set $f(0)=1$ without loss of generality.

Hence, in terms of the $N$-scaled rates $w_\pm(q)$, an arbitrary, symmetry preserving transformation reads $w_\pm(q)\to \tilde w_\pm(q) = f(q) w_\pm(q)$. The corresponding Hamiltonian $\ms{H}(q,p)$ transforms as 
\algn{
\ms{H}(q,p)\to\ms{\tilde H}(q,p)=f(q)\ms{H}(q,p)\,.
}
For small $q$, we may write 
\algn{
	f(q) \sim 1 + \zeta_2 q^2 + \ldots\,.
}
Applying the transformation to the general fourth-order Hamiltonian $\ms{H}$ in \Eqnref{hamun}, we observe that it only affects the coefficients $b_3$ and $b_4$ in the quartic Hamiltonian $\ms{H}_1$ according to
\algn{\eqnlab{btrans}
	b_3\to \tilde b_3 = \frac{a_1 \zeta_2}{2}\,,\qquad b_4 \to \tilde b_4 = b_4 + a_2 \zeta_2\,.
}
This transformation of $b_3$ and $b_4$ leaves $\sigma_0$ and $\omega_0$ unaffected. According to the discussion below \Eqnref{tceqn}, this ensures that the occurrence of the dynamical phase transition is robust against any such transformation. Note, however, that the transformation \eqnref{btrans} does in general change the higher-order terms $\propto \lambda_2$ and $\propto \lambda_3$ in $\ms{L}^{m,t}(m_0)$, and may thus affect the order of the dynamical phase transition.

The same conclusions can be drawn by considering the Riccati equation \eqnref{ric}. Transforming $\ms{\tilde H}(q,p)=f(q)\ms{H}(q,p)$ we observe that 
\algn{
\partial_q^2 \ms{\tilde H}(0,0)=\partial_q^2 \ms{H}(0,0)=0\,,\quad \partial_q\partial_p \ms{\tilde H}(0,0) = \partial_q\partial_p \ms{H}(0,0)=-2J(\beta_\text{c}-\beta_\text{q})/\tau\,,\quad \partial_p^2 \ms{\tilde H}(0,0)=\partial_p^2 \ms{H}(0,0)=4/\tau\,,
}
where we used $f(0)=1$ and $f'(0)=0$. Hence, \eqnref{ric} is invariant under arbitrary transformations of the rates that preserve the microscopic symmetries and the detailed balance condition.
\section{Strong fluctuations at critical point}
In this section, we obtain an explicit expression for the pre-factor $G(m,t)$ in $P(m,t) \propto G(m,t) \exp[-NV(m,t)]$ at $m=0$, which gives an indication for the presence of critical fluctuations around the optimal fluctuation $[q(s),p(s)]_{0\leq s\leq t}$ at the critical point. Upon substituting \Eqsref{limit1} and \eqnref{limit2} into \Eqnref{mastereqn2}, retaining terms of order $\ms{O}(N^{-1})$, we obtain
\begin{multline}\eqnlab{mastereqn3}
	-\partial_tV(m,t) =\sum_{\eta=\pm}w_\eta(m)\left\{\exp\left[ \eta2\partial_m V(m,t)	\right]-1\right\} + \frac1N\bigg(\beta_\text{q}J\sum_{\eta=\pm}w_\eta(m)\left\{\exp\left[ \eta2\partial_m V(m,t)	\right]-1\right\}\\ +2\sum_{\eta=\pm}w_\eta(m)\exp\left[ \eta2\partial_m V(m,t)	\right]\left\{-\eta\partial_m \log[w_\eta(m)]			-\partial^2_m V(m,t)\right\}\bigg)+ \ms{O}(N^{-2})\,.
\end{multline}
We now write $V(m,t) \sim \tilde V(m,t) + N^{-1}U(m,t)$ and evaluate the resulting equations at orders $\ms{O}(1)$ and $\ms{O}(N^{-1})$. The function $U(m,t)$ is related to the pre-factor $G(m,t)$, defined above, by $G(m,t) = \exp[-U(m,t)]$. Dropping the tilde $\tilde V\to V$, the $\ms{O}(1)$ equation associated with \Eqnref{mastereqn3} is the familiar Hamilton-Jacobi equation~\eqnref{hjeqn}. At order $\ms{O}(N^{-1})$, \Eqnref{mastereqn3} reads
\begin{multline}\eqnlab{v1order}
	0 = \partial_tU(m,t) 	+ \beta_\text{q}J\sum_{\eta=\pm}w_\eta(m)\left\{\exp\left[ \eta2\partial_m V(m,t)	\right]-1\right\} \\
	+ 2\sum_{\eta=\pm}w_\eta(m)\exp\left[ \eta2\partial_m V(m,t)	\right]\left\{\eta\partial_m U(m,t)-\eta\partial_m \log[w_\eta(m)]			-\partial^2_m V(m,t)\right\}\,.
\end{multline}
We now evaluate $U$ along the optimal fluctuation $q(t)$. Taking the total derivate of $U[q(t),t]$, we find
\algn{
	\dd{t}U[q(t),t] 	=& \partial_mU[q(t),t]\dot q(t) + \partial_t U[q(t),t]\,,\nn\\
					=&2\sum_{\eta=\pm}w_\eta[q(t)]\left\{\eta\partial_mU[q(t),t]\exp\left[ \eta2p(t)	\right]\right\} + \partial_t U[q(t),t]\,,
}
where we used the Hamilton equation for $q(t)$. This allows us to express \Eqnref{v1order} in terms of $\dd{t}U[q(t),t]$. We obtain
\algn{
	\dd{t}U[q(t),t] = 2\sum_{\eta=\pm}w_\eta[q(t)]\exp\left[ \eta2p(t)	\right]\left[\eta\partial_m \log\left\{w_\eta[q(t)]\right\}			+Z(t)\right] + \beta_\text{q}J\ms{H}[q(t),p(t)]\,.
}
Here, $Z(t) = \partial_m^2V[q(t),t]$ which follows the evolution equation~\eqnref{ricgen}. Upon integration from $0$ to $t$ we end up at
\algn{
	U[q(t),t] = U[q(0),0] +  \int_0^t\ed s\,\left\{2\sum_{\eta=\pm} w_\eta[q(s)]\exp\left[ \eta2p(s)	\right]\left[\beta_\text{q} J-\frac{1}{1-\eta q(s)}		+Z(s)\right] + \beta_\text{q}J\ms{H}[q(s),p(s)]\right\}\,.
}
Considering the particular solution $q(s)=p(s)=0$ and $Z(s) = z(s)$, relevant for $m=0$, this yields the explicit expression
\algn{
	U(0,t) 	=& U(0,0)+ \frac{1}{2}\log \left(\frac{\beta-\beta_\text{q}}{\beta_\text{c}-\beta_\text{q}}\right) + \frac{1}{2}\log \left[e^{-4(\beta_\text{c}-\beta_\text{q})J\left(\frac{t-t_\text{c}}{\tau }\right)}-1\right]\,,
}
and thus
\algn{
		G(0,t) 	=&\frac{G^\text{eq}(0)\sqrt{\beta_\text{c}-\beta_\text{q}}}{\sqrt{(\beta-\beta_\text{q})\left[e^{-4(\beta_\text{c}-\beta_\text{q})J\left(\frac{t-t_\text{c}}{\tau }\right)}-1\right]}}\,,
}
where $G^\text{eq}(0) = G(0,0)$ is the pre-factor that corresponds to the equilibrium distribution at $m=0$ prior to the quench. Close to the dynamical phase transition $(t_\text{c}-t)/\tau\ll1$, $G(0,t)$ diverges as
\algn{
	G(0,t) 	\sim&\frac{G^\text{eq}(0)}{\sqrt{4J(\beta-\beta_\text{q})}}\frac1{\sqrt{(t_\text{c}-t)/\tau}}\,,
}
indicating strong fluctuations around $[q(s),p(s)]_{0\leq s\leq t}$ at the critical point.